\documentstyle[prl,aps,psfig,twocolumn]{revtex}
\begin{document}
\twocolumn[\hsize\textwidth\columnwidth\hsize
\draft
%\onecolumn[\hsize\textwidth\columnwidth\hsize
           \csname @twocolumnfalse\endcsname
%\preprint{\parbox{5cm}{MPG--VT--UR 230/02}}
%\begin{titlepage}
\title{Phase diagram for rotating compact stars with two high density phases}
\author{D.~Blaschke \protect\cite{address1}}
\address{Institute for Nuclear Theory, University of Washington, Box 351550,
Seattle, WA 98195}
\author{H. Grigorian}
\address{Department of Physics, Yerevan State
University, 375025 Yerevan, Armenia}
\author{G. Poghosyan \protect\cite{address2}}
\address{Fachbereich Physik, Universit\"at Rostock, D-18051 Rostock, Germany}
\maketitle
\begin{abstract}
For the classification of rotating compact stars with two high density
phases a phase diagram in the angular velocity
($\Omega$) - baryon number ($N$) plane is investigated. The dividing
line $N_{\rm crit}(\Omega)$ between configurations with one and
two phases is correlated to a local maximum of the
moment of inertia and can thus be subject to experimental
verification by observation of the rotational behavior of
accreting compact stars. Another characteristic line, which also can
be measured is the transition line to black holes that of
the maximum mass configurations. The positions and
the shape of these lines are sensitive to changes in the equation of
state (EoS) of stellar matter. 
A comparison of the regional structure of phase diagrams   
is performed for polytropic and relativistic mean field type EoS and  
correlations between the topology of the
transition lines and the properties of two-phase EoS are obtained. 
Different scenarios of compact star evolution 
are discussed as trajectories in the phase diagram.
It is shown that a population gap in the $\Omega - N$ plane for
accreting compact stars would signal a high density phase transition 
and could be observed in the distribution of so called Z sources
of quasi periodic oscillations in low-mass X-ray binaries.

\end{abstract}
\vskip1cm]

At present, the existence of exotic phases of matter at 
high densities is under experimental investigation in ultrarelativistic 
heavy-ion collisions \cite{qm01} the most prominent being the
deconfined phase of QCD \cite{bkr}.
While the diagnostics of a phase transition in experiments with heavy-ion
beams faces the problems of strong nonequilibrium and finite size,
the dense matter in a compact star forms a macroscopic system in
thermal and chemical equilibrium for which effects signalling a
phase transition shall be most pronounced.

Signals of high density phase transitions have been suggested in the form 
of characteristic changes of observables such as the surface temperature
\cite{cool,bgvcool}, brightness \cite{magnetar}, pulse timing
\cite{frido} and rotational mode instabilities \cite{madsen}
during the evolution of the compact object. 
In particular the pulse timing signal has attracted much interest since it 
is due to changes in the kinematics of rotation. 
Thus it could be used not only to detect the occurrence, but also to 
determine the size of the high density matter core from the magnitude of 
the braking index deviation from the magnetic dipole value \cite{cgpb}. 
Besides of the isolated pulsars, one can consider also the accreting compact
stars in low-mass X-ray binaries (LMXBs) as objects from which we
can expect signals of a high density phase transition in their interior
\cite{fridonew,accmag}. 
The observation of quasiperiodic brightness oscillations (QPOs) 
\cite{klis} for some LMXBs has lead to very stringent constraints for masses 
and radii \cite{MLP} which according to \cite{strange} could even favour 
strange quark matter interiors over hadronic ones for these objects. 
Due to the mass accretion flow these systems are candidates for the formation
of the most massive compact stars from which we expect to observe
signals of the transition to either quark core stars, to a third
family of stars \cite{twins} or to black holes.

In this work we introduce a classification of rotating compact
stars in the plane of their angular frequency $\Omega$ and mass
(baryon number $N$) which we will call {\it phase diagram}. 
In this diagram, configurations with high density matter cores are separated
from conventional ones by a critical phase transition line.

Since the phase diagram of rotating compact objects seems to be a
more general approach for investigations of phase transition
effects in the interior of the star we assume that the
deconfinement transition could be a particular case besides of
other possibilities for phase transitions like pion or kaon
condensation as discussed, e.g. in \cite{migdal,reddy}. 
Therefore, our aim is to suggest it as a heuristic tool for obtaining
constraints on the EoS at high densities from the rotational
behaviour of compact stars. 
We will provide criteria under which a particular astrophysical scenario 
with spin evolution could be qualified to signal of high density phase 
transition.

%\section{Equation of State of Stellar matter with phase transition}

The true EoS that describes the interior of compact
stars is largely unknown. This results from the inability to
verify experimentally the different theories that describe the
strong interactions between baryons and the many-body theories of
dense matter at densities larger than about twice the nuclear
density \cite{sterglife}.

\begin{figure}
\psfig{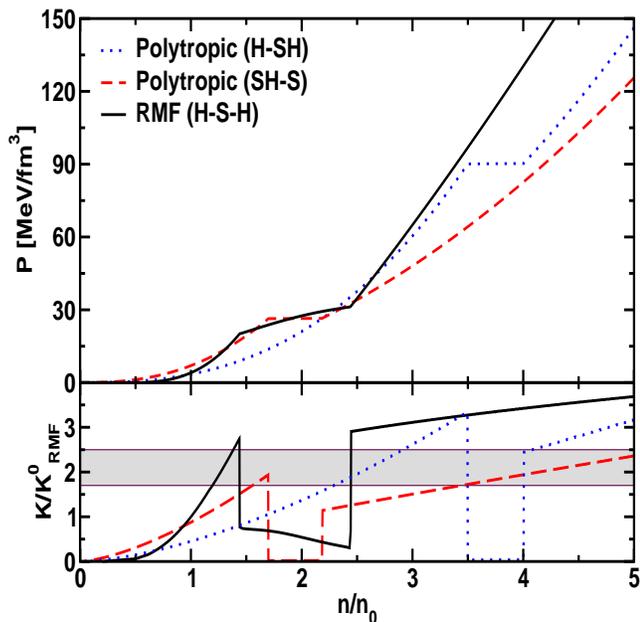}
\caption{Upper panel: Two polytropic type EoS models: SH-S (dashed line)
and H-SH (dotted line) are compared to a relativistic mean field
(RMF) EoS, which has a hard-soft-hard (H-S-H) phase transition between a
linear Walecka model for the low density hadronic phase and a dynamical quark
model for the high density quark matter phase with a soft mixed phase 
\protect\cite{cgpb}. 
SH-S ($\Gamma_L=2.5,~\Gamma_H=1.9$) represents a semihard-soft  
phase transition; H-SH ($\Gamma_L=2.6,~\Gamma_H=2.2$) hard-semihard one.
Lower panel: Incompressibilities for the EoS model of the upper panel 
(same line styles) in units of the value RMF in $\beta$- equilibrium,
$K^0_{RMF} = 1125$ MeV. The shaded region separates
the hard (H) and soft (S) parts of EoS and defines the semihardness (SH).} 
\label{EoSfig}
\end{figure}

Since our focus is on the elucidation of qualitative features of
signals from the high density phase transition in the pulsar timing we
will use a generic form of an equation of state (EoS) with such a
transition. 
We use the polytropic type  equation of state \cite{physrep}
in the form
\[
P_i=\frac{K^0_i~n_0}{\Gamma_i} \left(\frac{n}{n_0}\right)^{\Gamma_i};~~
\epsilon_i = \frac {P_i}{\Gamma_i -1} + m~n, \nonumber
\]
where $n$ is the baryon number density, $P_i$ - the pressure, and
$\epsilon_i$- energy density for both the low ($i=L$) and the high ($i=H$) 
density phases, respectively.   
$K^0_i=K_i(n_0)$ is the value of the incompressibility \cite{gbook}  
$K_i(n)=9~dP/dn$ at
the saturation density $n_0 = 0.17/fm^3$, 
$\Gamma_i= d \ln(P_i)/d \ln\epsilon_i$ the adiabatic index and 
$m$ is the nucleon mass. 
The phase transition between the lower and higher density phases is made
by the Maxwell construction \cite{hh} and compared to a relativistic mean 
field model consisting of a linear Walecka plus dynamical quark 
model EoS \cite{bbkr,bt00} with a Gibbs construction \cite{glen,cgpb}, 
see Fig.\ref{EoSfig}.

The quark matter part of this EoS is obtained from a dynamical
confining approach \cite{bbkr} in the generalization to three
flavors \cite{gbkg}, where the strange flavor remains confined at
the deconfinement transition for the light and appears only at
densities for which stars are close to the gravitational
instability. The difference to most of the models for quark
deconfinement in neutron star matter is that the ambiguity in the
choice of the bag constant for the quark matter phase can be
removed by a derivation of this quantity \cite{b+99,bt00} within
the dynamical confining approach \cite{bbkr,gbkg}. 

With the EoS models discussed above we have performed 
calculations of rotating compact star configurations 
assuming stationary, rigid rotation. 
For our treatment of rotation within general relativity we employ a
perturbation expansion following Refs. \cite{hartle}.
For a detailed discussion of the method and its application we
refer to \cite{cgpb}. The results of our calculations of rotating
compact star configurations can be classified in the plane of
angular velocity $\Omega$ and baryon number $N$ which we call 
{\it phase diagram}.

\begin{figure}[hbt]
\psfig{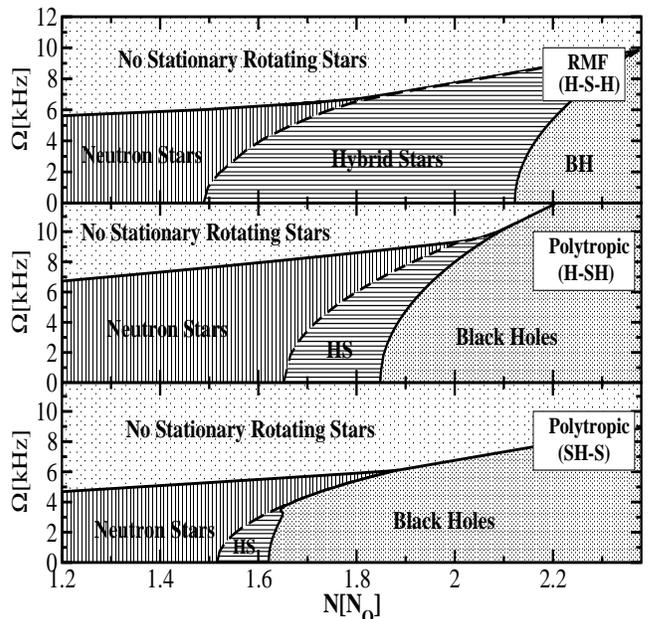}
\caption{Phase diagrams for the RMF EoS (upper panel) and two polytropic 
model EoS (lower panels). 
}
\label{phdiags}
\end{figure}

In Fig. \ref{phdiags} we display the phase diagrams for the rotating star 
configurations, which correspond to the three model EoS of Fig. \ref{EoSfig}.
These phase diagrams have four regions: 
(i) the region above the maximum frequency $\Omega >\Omega_{\rm K}(N)$ 
where no stationary rotating configurations are found, (ii) the region of
black holes $N > N_{\rm max}(\Omega)$, and the region of stable compact 
stars which is subdivided by the critical line $N_{\rm crit}(\Omega)$ into 
(iii) the region of hybrid stars for $N > N_{\rm crit}(\Omega)$ where 
configurations contain a core with a second, high density phase and 
(iv) the region of mono-phase stars
without such a core.

From the comparison of the regional structure of these three different phase
diagrams in Fig. \ref{phdiags} with the corresponding EoS
in Fig. \ref{EoSfig} we arrive at the main result of this paper that 
there are the following correlations between the topology
of the lines $N_{\rm max}(\Omega)$ and $N_{\rm crit}(\Omega)$ and the
properties of two-phase EoS: 
\begin{itemize}
\item[-] 
The hardness of the high density EoS determines the maximum mass of
the star, which is given by the line $N_{\rm max}(\Omega)$. 
Therefore  $N_{\rm max}(0)$ is proportional to the parameter $K_H(n_H)$, 
where $n_H$ is the density of the transition to high density phase.
\item[-] 
The onset of the phase transition line $N_{\rm crit}(0)$ depends on
the density $n_H$ and $K_L(n_L)$ where $n_L$ is the density of the transition
to the low density phase. 
\item[-] 
The curvature of the lines $N_{\rm max}(\Omega)$ and 
$N_{\rm crit}(\Omega)$ is proportional to the compressibility of the high and 
low density phases, respectively.
\end{itemize}

Therefore, a verification of the existence of the critical lines 
$N_{\rm crit}(\Omega)$ and $N_{\rm max}(\Omega)$ by observation of 
the rotational behavior of compact objects would
constrain the parameters of the EoS for neutron star matter.

\begin{figure}[htb]
\psfig{figure=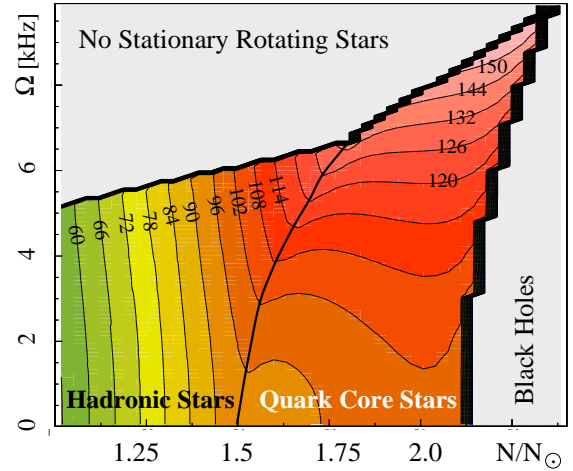,width=9.5cm,angle=0} \vspace{5mm}
\caption[]{Phase diagram for configurations of rotating compact
objects in the plane of angular velocity $\Omega$ and mass (baryon
number $N$). Contour lines show the values of the moment of
inertia in M$_\odot$km$^2$. The line $N_{\rm crit}(\Omega)$ which
separates hadronic from quark core stars corresponds the set of
configurations with a central density equal to the critical
density for the occurrence of a pure quark matter phase.
\label{phases}}
\end{figure}

The determination of $N_{\rm max}(\Omega)$, the border between compact stars
and black holes, is a traditional issue which has recently gained new impetus
due to the interpretation of recent LMXB data.
A new aspect characterizing the configurations with a phase transition is the 
critical line    $N_{\rm crit}(\Omega)$ which can be measurable by changes in 
the rotational dynamics since it is correlated with a local maximum of the 
moment of inertia $I(N,\Omega)$, the key quantity governing the rotational 
evolution via
\begin{equation}
\label{dyn}
\dot{\Omega}=
\frac{K(N,\Omega)}
{I(N,\Omega)+\Omega\left({\partial I(N,\Omega)}/{\partial \Omega}\right)_N}~.
\end{equation}
In Fig.\ref{phases} we show for the example of the RMF EoS with a
deconfinement phase transition from hadronic to quark matter how the 
structure of the lines of constant moment of inertia correlated with 
the critical lines in the phase diagram.

In Eq.(\ref{dyn}) $K=K_{\rm int}+K_{\rm ext}$ is the net 
torque acting on the
star due to internal and external forces.The internal torque is
given by $K_{\rm int}(N,\Omega)=-\Omega\dot N \left({\partial
I(N,\Omega)}/{\partial N}\right)_\Omega$~, the external one can be
subdivided into an accretion and a radiation term $K_{\rm
ext}=K_{\rm acc} + K_{\rm rad}$. The first one is due to all
processes which change the baryon number, $K_{\rm acc}=\dot
N~dJ/dN$ and the second one contains all processes which do not.
For the example of magnetic dipole and/or gravitational wave
radiation it can be described by a power law $K_{\rm
rad}=\beta\Omega^{n}$, see \cite{ghoshlamb,shapiro}.

To  prove that the appearance or disappearance of a high density phase
during the rotational evolution of the star could entail
observational consequences for the
angular velocity we consider three main representatives different 
classes of trajectories 
which could cross the critical line
on phase diagram.
These classes of tracks are: 
(a) spindown of isolated (non-accreting, $\dot N=0$)
pulsars due to magnetic dipole radiation \cite{frido,cgpb}, 
(b) spin up in accreting systems with weak magnetic field 
\cite{fridonew,accmag} ($N \simeq{\rm const}$, vertical tracks) and
(c) accretion either with strong magnetic field \cite{accmag} 
or for accreting binaries emitting gravitational
waves \cite{bildsten}, for which $\Omega \simeq {\rm const}$
(horizontal tracks).

In the case of (a) the spindown $K= K_{\rm rad}$ or (b) spinup  $K= K_{\rm
acc}$ evolutions (in both cases $K_{\rm int}<< K_{\rm ext}$)    
the  objects can undergo  a phase transition if the baryon number 
lies within the interval 
$N_{\rm crit}(\Omega = 0)< N < N_{\rm c}$, where $N_{\rm c}$ is the 
end piont of the critical line $N_{\rm crit}(\Omega)$.

If the core of compact star is soft enough (as in case (SH-S))
the critical line $N_{\rm crit}(\Omega)$ crosses $N_{\rm
max}(\Omega)$ at some $\Omega_c(N_c)< \Omega_K(N_c)$. 
This means, that massive mono-phase configurations with total baryon number
$N>N_c$ rotating with angular velocities in the interval 
$\Omega_c<\Omega < \Omega_K$ 
should encounter a transition to a black hole during the spin down evolution.

In cases, when the core EoS is harder (H-S-H and SH-H), the region of hybrid 
stars is a band which separates mono-phase configurations from black holes, 
see upper two panels of Fig. \ref{phdiags}. 

As it has been shown in \cite{cgpb} for the vertical tracks (a) and
(b) in the phase 
diagram, the braking index $n(\Omega)$ changes its value from $n(\Omega)>3$ 
in the region (iii) to $n(\Omega)<3$ in (iv). 
This is the braking index signal for a deconfinement transition 
introduced in Ref. \cite{frido}.

The third evolutionary track is accretion with strong magnetic fields 
\cite{accmag} and/or gravitational wave emission (horizontal tracks)
\cite{bildsten}.For this case the $\dot \Omega$ first 
decreases as long as the moment of
inertia monotonously increases with $N$. When passing the critical
line $N_{\rm crit}(\Omega)$ for the phase transition, the
moment of inertia starts decreasing and the internal torque term
$K_{\rm int}$ changes sign. This leads to a narrow dip for $\dot
\Omega (N)$ in the vicinity of this line. As a result, the phase
diagram gets overpopulated for $N \stackrel{<}{\sim} N_{\rm
crit}(\Omega)$ and depopulated for $N \stackrel{>}{\sim} N_{\rm
crit}(\Omega)$ up to the second maximum of $I(N, \Omega)$ close to
the black-hole line $N_{\rm max}(\Omega)$. 
This {\it population gap} marks the region of hybrid stars in the
phase diagram and is a measurable. Moreover, as we have seen 
population clustering of compact stars at the phase
transition line could be a signal for the
occurrence of stars with high density matter cores and a measure for
obtaining constraints on the EoS at high densities.
In contrast to this
scenario, in the case without a phase transition, the
moment of inertia could at best saturate before the transition to
the black hole region and consequently $\dot \Omega$ would also
saturate. This would entail a smooth population of the phase
diagram without a pronounced structure \cite{ssacc}.

As a strategy for the investigation of high density phase 
transitions in compact stars we suggest to perform a systematic 
observation of LMXBs for which the discovery of strong and remarkably
coherent high-frequency QPOs with the Rossi X-ray Timing Explorer 
has provided new information about the masses and rotation frequencies
of the central compact object \cite{klis,MLP,LM00}.
If, e.g., the recently discussed period clustering for Atoll- and Z-sources
\cite{fridonew,bildsten} will correspond to objects in a narrow
region of masses well below the maximum mass limit, this could be
interpreted as a signal for the
high density phase transition to be associated with a fragment of the
critical line in the phase diagram for rotating compact stars
\cite{accmag}. 

In the present work we have shown that the existence and
the shape of the suggested population gap for LMXBs in the phase
diagram will signal the occurence of a phase transition 
in the QCD EoS and constrains its properties
at high densities.

\section*{Acknowledgement}
We thank I. Bombaci, M. Colpi, N.K. Glendenning, A. Sedrakian and F. Weber
for their stimulating interest in our work,
C. Gocke, C.D. Roberts and S. Schmidt for their useful remarks.
D.B. thanks for partial support of the Department of
Energy during the program INT-01-2: "Neutron Stars" at the
University of Washington.
The work of G.P. was supported by DAAD fellowship A/01/08510 and by DFG
grant 436 ARM 17/7/00, that of H.G. by DFG grant 436 ARM 17/5/01.


\vspace{-5mm}
\begin{thebibliography}{}
\vspace{-13mm}
\bibitem[*]{address1}
{Permanent address: Fachbereich Physik,
Universit\"at Rostock, D-18051 Rostock, Germany}
\bibitem[\dagger]{address2}
{Permanent address: Department of Physics, Yerevan State
University, 375025 Yerevan, Armenia}
\bibitem{qm01}
T. J. Hallmann, D. E. Kharzeev, J. T. Mitchell, T. Ullrich (Eds.), 
{\it Quark Matter 2001: Proceedings},
Nucl. Phys. {\bf 698} (2002).
%%
\bibitem{bkr}
D. Blaschke, F. Karsch, C. D. Roberts (Eds.),
{\it Understanding Deconfinement in QCD}, World Scientific, Singapore (2000).
%
\bibitem{cool}
Ch. Schaab, B. Hermann, F. Weber, M. K. Weigel, Astrophys. J. {\bf 480}, L111
(1997).
%
\bibitem{bgvcool}
D. Blaschke, H. Grigorian, D. Voskresensky, Astron. Astrophys. {\bf
368}, 561 (2001).
%
\bibitem{magnetar}
A. Dar and A. DeR\'ujula, astro-ph/0002014.
%
\bibitem{frido}
N. K. Glendenning, S. Pei, F. Weber,
  Phys. Rev. Lett. {\bf 79}, 1603 (1997).
%
\bibitem{madsen}
J. Madsen, Phys. Rev. Lett. {\bf 85}, 10 (2000).
%
\bibitem{cgpb}
E. Chubarian, H. Grigorian, G. Poghosyan, D. Blaschke,
Astron. Astrophys. {\bf 357}, 968 (2000).
%
\bibitem{fridonew}
N. K. Glendenning, F. Weber, in {\it Physics of Neutron Star Interiors},
D. Blaschke, N. K. Glendenning, A. Sedrakian (Eds.), Springer Lect.
Notes Phys. {\bf 578}, 305 (2001).
%; [astro-ph/0003426].
%
\bibitem{accmag} G. Poghosyan, H. Grigorian, D. Blaschke,
Astrophys. J. {\bf 551}, L73 (2001); [astro-ph/0101002].
%
\bibitem{klis}
M. Van der Klis, Ann. Rev. Astron. Astrophys. {\bf 38}, 717 (2000).
%
\bibitem{MLP}
M. C. Miller, F. K. Lamb, D. Psaltis, Astrophys. J. {\bf 508}, 791 (1998).
%
\bibitem{strange}
X.-D. Li, I. Bombaci, M. Dey, J. Dey, E. P. J. van den Heuvel,
Phys. Rev. Lett. {\bf 83}, 3776 (1999).
%
\bibitem{twins}
N. K. Glendenning, Ch. Kettner, Astron. Astrophys. {\bf 353}, L9 (1999).
%
\bibitem{migdal}
A. B. Migdal, E. E. Saperstein, M. A. Troitsky, D. N. Voskresensky,
Phys. Rep. {\bf 192}, 179 (1990).
%
\bibitem{reddy}
J. A. Pons, S. Reddy, P. J. Ellis, M. Prakash, J. M. Lattimer,
Phys. Rev. {\bf C 62}, 035803 (2000).
%; T. Norsen, S. Reddy, Phys. Rev. {\bf C 63}, 065804 (2001).
%
\bibitem{sterglife}
N. Stergioulas, "Rotating Stars in Relativity", Living Rev.
Relativity 1, 8 (1998).
%
%\bibitem{poly}
%Tooper, R. F., ApJ 142, 1541 (1965) % ApJ Lett., 405, L29, (1965).
%
\bibitem{physrep}
G. E. Brown, Phys. Rep. {\bf 163}, 3 (1988).
%
\bibitem {hh}
H. Heiselberg, H. Hjorth-Jensen, Phys. Rev. Lett. {\bf 80}, 5485 (1999).
%
\bibitem{glen}
N. K. Glendenning, Phys. Rev. {\bf D 46}, 1274 (1992).
%
\bibitem{bbkr}
A. Bender, D. Blaschke, Yu. Kalinovsky, C. D. Roberts,
Phys. Rev. Lett. {\bf 77}, 3724 (1996).
%
\bibitem{gbkg}
C. Gocke, D. Blaschke, A. Khalatyan, H. Grigorian,
%{\it Equation of State for Strange Quark Matter in a Separable Model},
Preprint ECT*-2001-05; [hep-ph/0104183].
%
\bibitem{b+99}
D. Blaschke, H. Grigorian, G. Poghosyan, C.D. Roberts, S. Schmidt,
Phys. Lett. {\bf B 450}, 207 (1999).
%
\bibitem{bt00}
D. Blaschke, P.C. Tandy, in \protect\cite{bkr}, p. 218.
%
\bibitem{gbook}
N. K. Glendenning, {\it Compact Stars}, Springer, New York (2000).
%
\bibitem{hartle} J. B. Hartle, Astrophys. J. {\bf 150}, 1005 (1967);
J. B. Hartle, K. S. Thorne, Astrophys. J. {\bf 153}, 807 (1967);
%
%\bibitem{chubarian}
D. M. Sedrakian, E. V. Chubarian, Astrofizika {\bf 4}, 239; 551 (1968).
%
\bibitem{ghoshlamb}
P. Ghosh, F. K. Lamb, Astrophys. J. {\bf 234}, 296 (1979).
%
\bibitem{shapiro}
S. L. Shapiro, S. A. Teukolsky,
{\it Black Holes, White Dwarfs, and Neutron Stars},
Wiley, New York (1983).
%
\bibitem{bildsten} L. Bildsten, Astrophys. J. {\bf 501}, L89 (1998).
%
\bibitem{LM00}
F. K. Lamb, M. C. Miller, HEAD (2000); [astro-ph/0007460].
%
\bibitem{ssacc}
D. Blaschke, I. Bombaci, H. Grigorian, G. Poghosyan,
New Astronomy {\bf 7}, 107 (2002).
%{\it Timing evolution of accreting strange stars}
%[astro-ph/0110443].
\end{thebibliography}
\end{document}